\documentclass[conference]{IEEEtran}
\IEEEoverridecommandlockouts
\usepackage{cite}
\usepackage[numbers, sort&compress]{ natbib}
\usepackage[utf8]{ inputenc} 
\usepackage[T1]{fontenc}
\usepackage{url}
\usepackage{ ifthen}
\usepackage[cmex10]{amsmath}
\usepackage{array} 

\usepackage{enumerate}
\usepackage{amssymb}
\usepackage{amsmath} 
\usepackage[lined,boxed,commentsnumbered, ruled]{algorithm2e}
\usepackage{mathrsfs}
\usepackage{bm}
\usepackage{float}
\usepackage{tikz}
\usepackage{algorithmic}
\usetikzlibrary{arrows}
\usepackage{subfigure}
\usepackage{graphicx,booktabs,multirow}
\usepackage{epstopdf}
\usepackage{subfigure}
\usepackage{multirow}
\usepackage{tabularx} 
\usepackage{booktabs}
\usepackage{stmaryrd}
\usepackage{tikz}
\usepackage{pgfplots}
\usepackage{textcomp}
\usepackage{siunitx}
\pgfplotsset{compat=1.12}

\definecolor{colorhkust}{RGB}{20,43,140}
\definecolor{colortsinghua}{RGB}{116,52,129}
\definecolor{color1}{RGB}{128,0,0}

\usepackage{amsthm}

\theoremstyle{definition}

\renewcommand{\baselinestretch}{0.94}

\def\BibTeX{{\rm B\kern-.05em{\sc i\kern-.025em b}\kern-.08em
		T\kern-.1667em\lower.7ex\hbox{E}\kern-.125emX}}
\begin{document}

\title{Over-the-Air Computation via Cloud Radio Access Networks \thanks{This work is supported in part by the National Natural Science Foundation of China (NSFC) under Grant 62001294.}}

%

\author{\IEEEauthorblockN{$\text{Lukuan Xing}^{*\dag\ddag}$, $\text{Yong Zhou}^{*}$, and $\text{Yuanming Shi}^{*}$}
\IEEEauthorblockA{${}^{*}$\text{School of Information Science and Technology, ShanghaiTech University, Shanghai $201210$, China} \\
${}^{\dag}$\text{Shanghai Institute of Microsystem and Information Technology, Chinese Academy of Sciences, China}\\
${}^{\ddag}$\text{University of Chinese Academy of Sciences, Beijing $100049$, China}\\
E-mail: \{xinglk, zhouyong, shiym\}@shanghaitech.edu.cn}
}

\maketitle
\begin{abstract}
Over-the-air computation (AirComp) has recently been recognized as a promising scheme for a fusion center to achieve fast distributed data aggregation in wireless networks via exploiting the superposition property of multiple-access channels. Since it is challenging to provide reliable data aggregation for a large number of devices using AirComp, in this paper, we propose to enable AirComp via the cloud radio access network (Cloud-RAN) architecture, where a large number of antennas are deployed at separate sites called remote radio heads (RRHs). However, the potential densification gain provided by Cloud-RAN is generally bottlenecked by the limited capacity of the fronthaul links connecting the RRHs and the fusion center. To this end, we formulate a joint design problem for AirComp transceivers and quantization bits allocation and propose an efficient algorithm to tackle this problem. Our numerical results shows the advantages of the proposed architecture compared with the state-of-the-art solutions.

%

\end{abstract}

\setlength{\abovedisplayshortskip}{3.5pt}
\setlength{\belowdisplayshortskip}{3.5pt}

\section{Introduction}
It is expected that a huge number of Internet of things (IoT) devices will be connected to wireless networks to boost the proliferation of intelligent services in our daily life \cite{8808168}, \cite{2015Large}, \cite{2020Communication}. To realize this promising vision, one key challenge is the urgent need of ultra-fast wireless data aggregation, which pervades a wide range of applications, including massive machine type communication \cite{7937902} and on-device federated machine learning \cite{8952884}. As a result, we need to use wireless communication technology to quickly collect and process data distributed across a massive number of devices. However, it is difficult to use the conventional interference-avoiding channel access schemes to aggregate massive distributed data while achieving high spectrum utilization efficiency and low network latency \cite{2018MIMO,2020Optimized}. To overcome this challenge, a promising solution called over-the-air computation (AirComp) has recently emerged as a multi-access scheme that exploits the superposition property of multiple-access channels \cite{2007Computation}, \cite{6557530}, \cite{2018A}, \cite{2018Blind}. 

There have been extensive research works on investigating the different e{\color{black} network architectures for AirComp} \cite{2018MIMO}, \cite{2020Optimized}, \cite{2007Computation}, \cite{2019Over}, \cite{2020interference}. In particular, {\color{black} a single-antenna AirComp system} was developed in \cite{2020Optimized} to minimize the computation error by jointly optimizing the transmit power at devices and a signal scaling factor at the base station (BS) by using channel-inversion power control method.	 {\color{black} For multiple-input multiple-output (MIMO) AirComp systems}, zero-forcing precoding was designed at the transmitter and the multi-antenna server attempts to apply receive beamforming, called aggregation beamforming, to achieve simultaneous magnitude alignment of spatially multiplexed multiuser signals to receive parallel functional streams \cite{2020aircompmag}. Further, {\color{black} intelligent reflecting surface (IRS) aided AirComp system} was developed in \cite{AirCompIRS}, \cite{IRSMag}, \cite{9163314}, \cite{federatedIRS} to build controllable wireless environments, thereby boosting the received signal power significantly by optimizing the  phase shifts matrix at IRS.

Despite of the previous research, alleviating {\color{black} the performance deterioration due to channel fading} is still a great challenge for the existing AirComp systems \cite{2020aircompmag}. {\color{black} The conventional massive MIMO AirComp system can not receive reliable signals transmitted by the devices which are far away from the BS due to the severe path loss.} Hence, we propose to deploy Cloud-RAN architecture to support AirComp, in which the devices upload local data to the baseband unit (BBU) through distributed RRHs and the signal is transmitted through fronthaul links between RRHs and BBU \cite{6786060}. {\color{black} Relying on a large number of RRHs geographically spread out over a region densely, {\color{black} Cloud-RAN facilitates scaling and increasing the baseband processing density and reducing path loss of the channels between the RRHs and devices.} Furthermore, it also achieves a high diversity gain against channel fading by exploiting the independent fading of their signals to aggregate data more accurately in AirComp systems.} However, in practice, the capacity of fronthaul links connecting the RRHs and the BBU in Cloud-RAN is limited, which limits the performance gain provided by dense RRHs \cite{8452186}, \cite{7208841}, \cite{7134796}.

To tackle the challenge due to limited fronthaul capacity, we formulate an optimization problem of joint devices' transmit beamforming and RRHs' quantization bits allocation and BBU's receive beamforming design to minimize the mean square error (MSE) of AirComp. We propose an efficient solution to this complicate optimization problem. Specifically, the BBU jointly optimizes the transceivers and quantization bits allocation to minimize the MSE based on the alternating optimization technique. Furthermore, {\color{black} numerical results demonstrate that our proposed solution enjoys the near optimal performance, and Cloud-RAN architecture for AirComp outperforms the conventional massive MIMO architecture}.

The rest of this paper is organized as follows. In Section II, we describe our system model of Cloud-RAN architecture for AirComp and formulate the MSE minimization problem. In Section III, we elaborate on our proposed optimization approach to the formulated problem. Section IV provides the simulation results to verify the effectiveness of Cloud-RAN architecture for AirComp. Finally, we conclude our work in Section V.

\section{System Model and Problem Formulation}

\subsection{System Model}

\begin{figure}[t]
	\centering 
	\includegraphics[width=8cm]{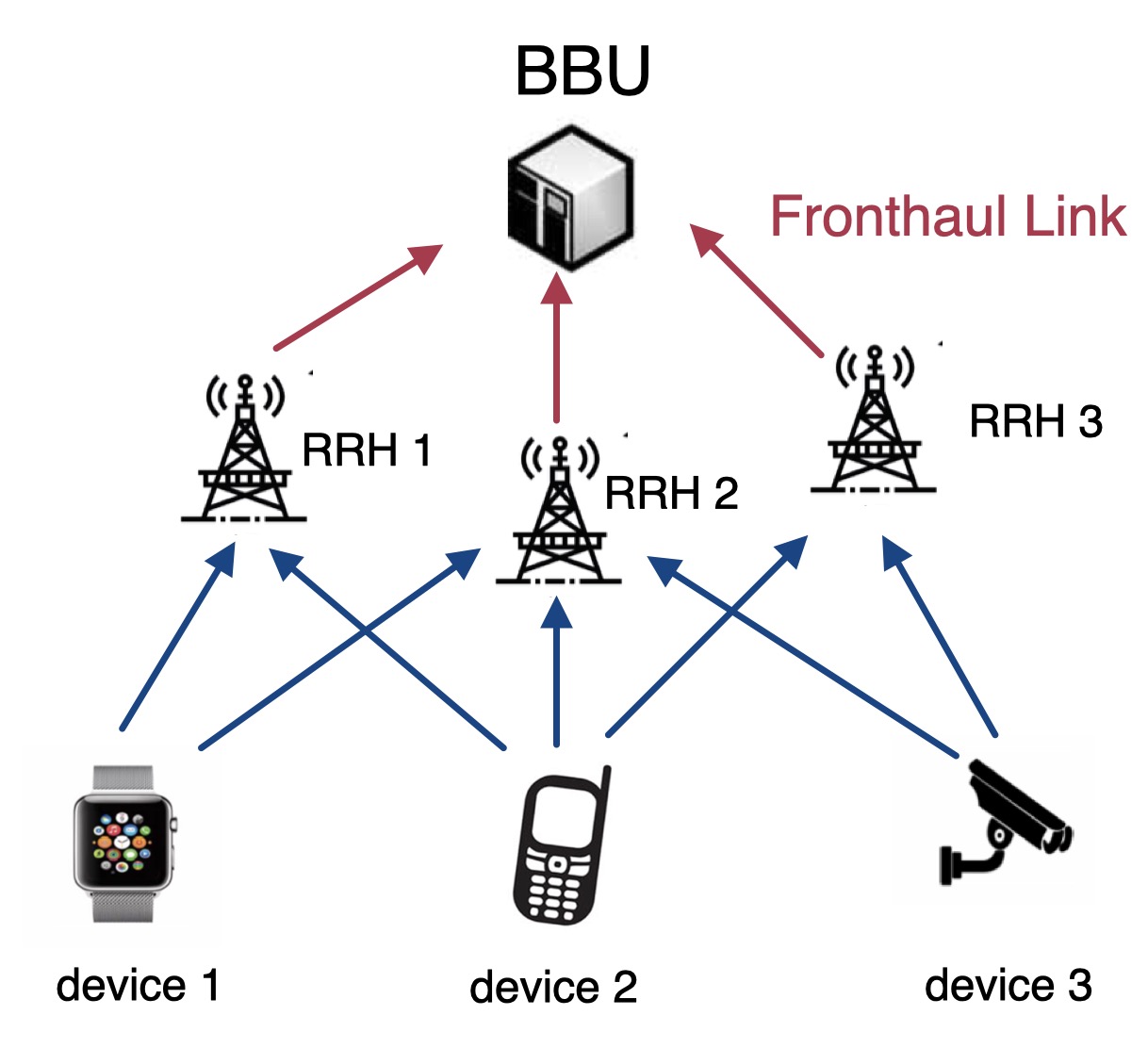}
	\caption{Cloud-RAN system}
	\label{system model}
\end{figure}

As illustrated in Fig. \ref{system model}, we consider an AirComp task performed on a multi-antenna Cloud-RAN system which consists of $N_W$ single-antenna devices, $N_A$ multi-antenna RRHs and one BBU. Each RRH is equipped with $M$ antennas. Each device sends their local data to the BBU through $N_A$ multi-antenna RRHs, while the RRHs transmit the information to the BBU via a fronthaul link which is modeled as a digital link of capacity $C$ bits/sample. We define the sets $\mathcal{N}_W = \{1, 2,\cdots, N_W \}$ and $\mathcal{N}_A = \{1, 2,\cdots, N_A\} $ for the indices of devices and RRHs, respectively.

For a specific time slot $t$, we define the data aggregated at device $k$ as $\theta_k^{(t)} \in \mathbb{C}$. Then the target function for aggregating local updates at the BBU can be written as
\begin{eqnarray}
f{(t)} = \Phi\left( \sum_{k\in \mathcal{N}_W} \phi_k(\theta_k^{(t)}) \right),
\end{eqnarray}
where $\Phi$(·) is the post-processing function at BBU, and $\phi_k$ is the pre-processing function of device $k$. We denote $x_k^{(t)} := \phi_k(\theta_k^{(t)})$ as the transmit symbols. In this work, we assume that transmit symbols are normalized to have unit variance.

In order to estimate the target function at time slot $t$, the BBU aims to recover the variable which can be given in the form as
\begin{eqnarray}\label{target}
g^{(t)} := \sum_{k\in \mathcal{N}_W} x_k^{(t)}.
\end{eqnarray} 

To simplify the notation, we omit the time slot index by writing $g$ and $x_k$ instead of $g^{(t)}$ and $x_k^{(t)}$. Then we consider the case with quasi-static flat-fading channels, where the channel conditions remain unchanged in a certain time slot but may vary from one to another. The received parameter at RRH $i$ can be written as
\begin{eqnarray}
\mathbf{y}_i = \sum_{k\in \mathcal{N}_W} \mathbf{h}_{i,k}b_kx_k + \mathbf{z}_i ,\quad i\in \mathcal{N}_A
\end{eqnarray} 
where $\mathbf{y}_i = [y_{i,1},\ldots,y_{i,M}]^\mathsf{T}$ with $y_{i,m},1\leq m \leq M$ denoting the signal received at the $m$-th antenna of RRH $i$, and $b_k$ denotes the transmitter scalar of device $k$, $\mathbf{h}_{i,k} \in \mathbb{C}^M$ denotes the communication channel vector from the device $k$ to the RRH $i$ and the interference channel vector between the radar transmitter and the BS receiver, and $\mathbf{z}_i \sim \mathcal{CN}(0,\sigma^2_z \mathbf{I})$ denotes the additive white Gaussian noise at RRH $i$. We assume that $\{\mathbf{z}_i\}_{i\in\mathcal{N}_A}$ are independent over $i$.

Furthermore, in practice, each device $k\in \mathcal{N}_W$ is constrained by a power budget $P_0$, i.e., the transmit power constraints of each device can be written as
\begin{eqnarray}\label{power constrain}
\mathbb{E}(|b_kx_k|^2) = |b_k|^2 \leq P_k.
\end{eqnarray}

Each multi-antenna RRH quantifies and forwards the received baseband symbol $\mathbf{y}_i$ to the BBU via the limited-capacity fronthaul link. Each RRH first demodulates the signal received from each antenna to the baseband, and then conducts a scalar quantization over each output in parallel, and finally forwards the quantized bits to the BBU via the fronthaul link. Specifically, we apply the uniform quantization to each element of compressed signal $\bar{\mathbf{y}_i}=[\bar y_{i,1},\ldots,\bar y_{i,L}]^\mathsf{T}$ at RRH $i$ via separate I/Q scalar quantization and the baseband quantized signal can be given as
\begin{eqnarray}
\mathbf{\hat y}_i = \mathbf{y}_i + \mathbf{q}_i = \sum_{k\in \mathcal{N}_W} \mathbf{h}_{i,k}b_kx_k + \mathbf{z}_i + \mathbf{q}_i, \quad i\in \mathcal{N}_A
\end{eqnarray}
where $\mathbf{q}_i = [q_{i,1},\ldots,q_{i,M}]^\mathsf{T}$ with $q_{i,m}, 1\leq m \leq M$, modeling the quantization distortion for the $y_{i,m}$ as being independent of $y_{i,m}$ and distributed as $q_{i,m} \sim \mathcal{CN}(0,\omega_{i,m})$. Let $C_{i,m}$ denote the number of bits that RRH $i$ uses to quantize the I-branch or Q-branch of $\bar y_{i,m}$. According to the results of \cite{7208841}, the quantization noise level $\omega_{i,m}$ due to $q_{i,m}$ for uniform quantization can be written as
\begin{eqnarray}\label{quantization noise element}
\omega_{i,m} = 3\left(\sum_{k\in \mathcal{N}_W}| \mathbf{e}_m^\mathsf{T} \mathbf{h}_{i,k}|^2|b_k|^2+\sigma^2_{z}\right)2^{-2C_{i,m}},
\end{eqnarray}
where $\mathbf{e}_m$ denotes the unit vector whose $m$-th entry is 1.

Note that $q_{i,m}$'s are independent over $m$ due to independent scalar quantization for each element of ${\mathbf{y}_n}$, and also over $i$ due to independent processing at different RRHs. Thus the covariance matrix of $\mathbf{q}_i$ is a function of $b_k,k\in \mathcal{N}_W$ as well as $C_{i,m},i\in \mathcal{N}_A$, which is given by
\begin{eqnarray}
\mathbf{q}_i = diag(\omega_{i,1},\ldots,\omega_{i,M})
\end{eqnarray}
Then, each RRH forwards the quantized bits to the BBU via the fronthaul link. The transmission rate of RRH $i$'s fronthaul link is expressed as
\begin{eqnarray}
	T_i = 2B\sum^M_{m=1}C_{i,m}, \quad i\in \mathcal{N}_A
\end{eqnarray}where $B$ is the channel bandwidth.

Based on the received quantized signals $\mathbf{\hat y}_i,\forall i$, the BBU estimates the target function $g$ in (\ref{target}). To elaborate, we define a vector $\mathbf{\hat y} = [\mathbf{\hat y_1}^\mathsf{T},\mathbf{\hat y_2}^\mathsf{T},\ldots,\mathbf{\hat y_{N_A}}^\mathsf{T}]$ which is stacked by the quantized signal at each RRH, and denote $\mathbf{\mathbf{h}}_k = [\mathbf{h}_{1,k}^\mathsf{T},\mathbf{h}_{2,k}^\mathsf{T},\cdots,\mathbf{h}_{N_A,k}^\mathsf{T}]^\mathsf{T}$ as the channel vector stacked by the channel vector from device $k$ to RRHs. Then, vector $\mathbf{\hat y}$ can be written as
\begin{eqnarray}
\mathbf{\hat y} = \sum_{k\in \mathcal{N}_W}\mathbf{h}_k b_k x_k + \mathbf{z+q},
\end{eqnarray}
where we define $\mathbf{z} = [\mathbf{z}_1^\mathsf{T},\ldots,\mathbf{z}_{N_A}^\mathsf{T}]^\mathsf{T} \sim \mathcal{CN}(\mathbf{0,\sigma^2_zI})$ and $\mathbf{q}=[\mathbf{q}_1^\mathsf{T},\ldots,\mathbf{q}_{N_A}^\mathsf{T}]^\mathsf{T} \sim \mathcal{CN}(\mathbf{0,\Omega)}$ with $\mathbf{\Omega} = diag(\mathbf{q}_1,\ldots,\mathbf{q}_{N_A}) $ .

By assuming that BBU performs a linear estimation of the target parameter $g$ from $\hat y$, the estimation of $g$ at the BBU can be given by
\begin{eqnarray}
\begin{aligned}
\hat g 
&= \mathbf{m}^\mathsf{H}\hat{\mathbf{y}} \\
&= \mathbf{m}^\mathsf{H} \sum_{k\in \mathcal{N}_W}\mathbf{h}_k b_k x_k + \mathbf{m}^\mathsf{H}\mathbf{(z+q)},
\end{aligned}
\end{eqnarray}
where $\mathbf{m} \in \mathbb{C}^{N_AL\times 1}$ is the receiver beamforming vector. Then each element of the target vector $g$ can be obtained at the BBU through (\ref{target}).

The distortion of $\hat g$ with respect to the target value $g$ can be measured by the mean-squared-error ($\bf{MSE}$) which is given as
\begin{eqnarray}\label{mse}
\begin{aligned}
\bf{MSE}&(\hat g,g)
= \mathbb{E}(|\hat g - g|^2) \\
&= \sum_{k \in \mathcal{N}_W} |\mathbf{m}^\mathsf{H} \mathbf{h}_k b_k -1|^2 + \mathbf{m}^\mathsf{H}(\sigma_z^2\mathbf{I} + \mathbf{\Omega})\mathbf{m}.
\end{aligned}
\end{eqnarray}

\subsection{Problem Formulation}
In this paper, our objective is to minimize $\bf{MSE}$ that quantifies the distortion after the decoding process at the BBU by optimizing the devices' transmit beamforming $\{b_k\},\forall k$, the receive beamforming vector at the BBU $\mathbf{m}$, as well as the quantization bits allocation at each RRH $\{C_{i,m}\},\forall i,m$. Specifically, the formulated optimization problem can be expressed as:
\begin{eqnarray}\label{problem 1}
\mathop{\text{minimize}}_{\{\mathbf{b}_k\},\{C_{i,m}\},\mathbf{m}}
&&\bf{MSE}(\hat g,g) \\
\text{Subject to}
&& |b_k|^2 \leq \bar P_k,  \forall k \\
&& 2B\sum^M_{m=1}C_{i,m} \leq \bar T_i,\forall i,m\\
&& C_{i,m} \in \mathbb{N}^{+},\forall i,m .
\end{eqnarray}

It can be observed that the optimization problem (\ref{problem 1}) is a non-convex problem since all the optimization variables are coupled in the objective function, and the quantization bits at each RRH $\{C_{i,m}\},\forall i,m$ are discontinuous variables. In Section III, we shall leverage the alternating optimization method to solve this problem.

\section{Optimization Framework}
In this section, we propose to solve problem (\ref{problem 1}) by utilizing the alternating optimization approach. Specifically, the receive beamforming vector $\mathbf m$, the devices' transmit beamforming $\{b_k\},\forall k$ and the quantization bits allocation $\{C_{i,m}\},\forall i,m$ are optimized in an alternative manner until the algorithm converges.

\subsection{Optimizing transmit beamforming and Receive Beamforming}
We firstly fix the quantization bits allocation in problem (\ref{problem 1}) to optimize transmit power control and receive beamforming vector by solving the  following problem:
\begin{eqnarray}\label{problem 3}
\mathop{\text{minimize}}_{\{\mathbf{b}_k\},\mathbf{m}}
&&{\bf{MSE}} (\hat g,g) \\
\text{Subject to}
&& |b_k|^2 \leq \bar P_k, \quad \forall k .
\end{eqnarray}
Problem (\ref{problem 3}) is still a non-convex problem since transmit beamforming $\{b_k\},\forall k$ and receive beamforming vector $\mathbf m$ are coupled. However, either fix receive beamforming vector $\mathbf{m}$ or fix transmit power constraints $\{b_k\}$ can reduce the problem to be convex. Hence, with fixed $\mathbf{m}$, we efficiently solve the following problem by applying interior-point method \cite{boyd2004convex}:
\begin{eqnarray}\label{problem 3a}
\mathop{\text{minimize}}_{\{\mathbf{b}_k\}}
&&{\bf{MSE}} (\hat g,g) \\
\text{Subject to}
&& |b_k|^2 \leq \bar P_k, \quad \forall k .
\end{eqnarray}
Let $\mathbf{\bar b} = [\bar b_1,\ldots, \bar b_k]^\mathsf{T}$ denote the solution to problem (\ref{problem 3a}). Note that given a certain transmit beamforming, finding the optimal receive beamforming vector $m$ becomes a quadratic optimization problem without any constraint. The closed-form solution is given as
\begin{eqnarray}\label{obtain m}
	\mathbf{\bar m} = \left( \sum_{k \in \mathcal{N}_W} |\bar b_k|^2 \mathbf{h}_k \mathbf{h}_k^\mathsf{T} + \sigma_z^2 \mathbf I + \Omega \right)^{-1} \sum_{k \in \mathcal{N}_W} \bar b_k \mathbf{h}_k.
\end{eqnarray} 

\begin{algorithm}[t] \label{overall algorithm}
 \caption{Overall Algorithm for Solving Problem (\ref{problem 1})}
 \begin{algorithmic}[1]
  \STATE Initialize: Set $C_{i,m}^{(0)} = \lfloor{\bar T_i/(2BM)}\rfloor,\forall i,m,$ and $i=0$. 

  \REPEAT 
  \STATE i = i + 1;
  \STATE Solve sub-problem (\ref{problem 3}) and obtain $\{b_k^{(i)}\}$ and by\\ using interior-point method with $\bar C_{i,m}^{(i)} = C_{i,m}^{(i-1)}$, $\forall i,m$;
  \STATE Substitute $\{b_k^{(i)}\}$ into (\ref{obtain m}) to obtain $\mathbf{m}^{(i)}$;
  \STATE Obtain continuous $\{\bar C_{i,m}^{(i)}\}$ by solving sub-problem (\ref{problem 4}$^*$) with inter-perior method.
  \STATE Apply bisection method to $\{\bar C_{i,m}^{(i)}\}$: \\ \quad Initialize $\tau_{i}^{min} = 0, \tau_{i}^{max} = 1$,$\forall i$\\ \quad For $i=1:N$, repeat \\ \qquad (i). $\tau_i = (\tau_{i}^{min} + \tau_{i}^{max})/2$; \\ \qquad (ii). Substitute $\tau_i$ into (\ref{quantize bits}). If $\{\hat C_{i,m}^{(i)}\}$ satisfy \\ \qquad \qquad $2B\sum_{m=1}^M \leq \bar T_n$, set $\tau_i^{max} = \tau_i$; \\ \qquad \qquad otherwise, set $\tau_i^{min} = \tau_i$; \\ \quad Until $\tau_i^{max} - \tau_i^{min} \leq \epsilon_1$, where $\epsilon_1$ is the\\ \quad arithmetic accuracy of bi-section method.
  \STATE Update the solution of problem (\ref{problem 1}).
  \UNTIL \\ 
  	MSE$^{(i)} - $MSE$^{(i-1)} \leq \epsilon_2$, where MSE$^{(i)}$ is the\\ objective value of problem (\ref{problem 1}) achieved by $\mathbf{m^{(i)}}$,\\ $\{b_k^{(i)}\}$ and $\{C_{i,m}^{(i)}\}$, and $\epsilon_2$ is the arithmetic accuracy\\ of the overall algorithm.
  
 \end{algorithmic}
\end{algorithm}

\subsection{Optimizing Quantization Bits Allocation}

In this subsection, we fix transmit beamforming and receive beamforming vector and optimize quantization bits allocation by solving the following problem:
\begin{eqnarray}\label{problem 4}
\mathop{\text{minimize}}_{\{C_{i,m}\}}
&&{\bf{MSE}} (\hat g,g)' \\
\text{Subject to}
&& 2B\sum^M_{m=1}C_{i,m} \leq \bar T_i,\quad \forall i,m\\
&& C_{i,m} \in \mathbb{N}^{+},\forall i,m
\end{eqnarray}
where $\bf{MSE}(\hat g,g)'$ is obtained by substituting $\mathbf{\bar b}$ and $\mathbf{\bar m}$ into (\ref{mse}), and can be written as
\begin{eqnarray}
\begin{aligned}
{\bf{MSE}} &(\hat g,g)' =\\
 &\sum_{k \in \mathcal{N}_W} |\mathbf{\bar m}^\mathsf{H} \mathbf{h_i}\bar b_k -1|^2 + \mathbf{\bar m}^\mathsf{H}(\sigma_z^2\mathbf{I} + \mathbf{\bar \Omega})\mathbf{\bar m},
\end{aligned}
\end{eqnarray}
where $ \bar{\mathbf{\Omega}}$ is obtained by substituting $\{ \bar b_k \}$ into (\ref{quantization noise element}).

Problem (\ref{problem 4}) is challenging to be solved due to the integer constraints for quantization bits $C_{i,m}$. Note that if quantization bits allocation $C_{i,m},\forall i,m$ is assumed to be continuous, the quantization noise power $\omega_{i,m}$ will turn to be a continuous function. In the following, we first solve the relaxation of problem (\ref{problem 4}) without integer constraints which we denote as problem (\ref{problem 4}$^*$). Then we propose an efficient algorithm to obtain a set of integer solutions for all $\{C_{i,m}\},\forall i,m$ based on the solution of problem (\ref{problem 4}$^*$). Firstly, we have
\begin{eqnarray}
\begin{aligned}
\mathbf{m}^\mathsf{H}\mathbf{\Omega m} 
& =\sum_{i=1}^{N_A}\sum_{m=1}^{M}\omega_{i,m}|\bar m_{(i-1)M+m}|^2 \\
& =\sum_{i=1}^{N_A}\sum_{m=1}^{M}\xi_{i,m}2^{-2C_{i,m}},
\end{aligned}
\end{eqnarray}
where $\bar m_{j}$ denotes the $j$-th element of $\mathbf{\bar m}$, $1\leq j\leq N_A\times M$ and 
\begin{eqnarray}
\xi_{i,m} = 3|\bar m_{(i-1)M+m}|^2 \left( \sum_{j=1}^{N_W}b_j|\mathbf{h}_{i,j}|^2 + \sigma_z^2 \right),\forall i,m
\end{eqnarray}
Note that $\xi_{i,m}$ can be interpreted as the effective quantization noise power due to the quantized dimension at RRH $i$. Further, it is worth noting that both the first term of $\bf{MSE}(\hat g,g)$ i.e. $\sum_{k \in \mathcal{N}_W} |\mathbf{\bar m}^\mathsf{H} \mathbf{h_i}\bar b_k -1|^2$ and $\sigma_z^2\mathbf{I}$ remains constant in this alternating optimization step, we can reformulate problem (\ref{problem 4}$^*$) as the following optimization problem.
\begin{eqnarray}\label{problem 5}
\mathop{\text{minimize}}_{\{C_{i,m}\}}
&&\sum_{i=1}^{N_A}\sum_{m=1}^{M}\xi_{i,m}2^{-2C_{i,m}} \\
\text{Subject to}
&& 2B\sum^M_{m=1}C_{i,m} \leq \bar T_i,\quad \forall i,m
\end{eqnarray}
Problem (\ref{problem 5}) can be shown to be a convex problem and thus we can solve it via the interior-point method. Let ${C_{i,m}}$ denote the solution of Problem (\ref{problem 5}).
Considering that the solution of problem (\ref{problem 4}$^*$) may not satisfy all the integer constraints, inspired by \cite{7134796}, we propose an efficient algorithm to obtain a set of integer solutions. In the following, we round each $C_{i,m},\forall i,m$ to its nearby integer as follows.
\begin{eqnarray}\label{quantize bits}
\hat C_{i,m} = \left\{
\begin{aligned}
\text{floor}(\bar C_{i,m}),&\quad \text{if } \bar C_{i,m} - \text{floor}(\bar C_{i,m}) \leq \tau_i,  \\
\text{Ceil}(\bar C_{i,m}),&\quad \text{otherwise},
\end{aligned}
\right.
\end{eqnarray}
where $0\leq\tau_i\leq1,\forall i$.

It is worth noting that we can always find a feasible solution of $\{\hat C_{i,m}\}, \forall i,m$ by simply setting $\tau_i = 1,\forall i$ in (\ref{quantize bits}). Next, we show how to optimize $\{\tau_i\}, \forall i$ to find a better feasible solution. {\color{black} Since $\{\hat C_{i,m}\},\forall i,m$ increases as $\tau_i,\forall i$ becomes smaller, the resulting $\{\hat C_{i,m}\},\forall i,m$ from (\ref{quantize bits}) achieves lower MSE, while the individual quantization bits constraints for RRHs become more difficult to be satisfied.} Thus, we propose to utilize the bisection method to find the optimal values of $\{\hat C_{i, m}\},\forall i,m$ and substitute it into (\ref{quantize bits}) to obtain $\{\hat C_{i, m}\},\forall i,m$. {\color{black} The bisection method is specified in the Step 7 of Algorithm \ref{overall algorithm}}.


The proposed algorithm for solving problem (\ref{problem 1}) is summarized in Algorithm \ref{overall algorithm}.

\begin{figure}[t]
	\begin{subfigure} 
	\centering 
	\includegraphics[width=9cm]{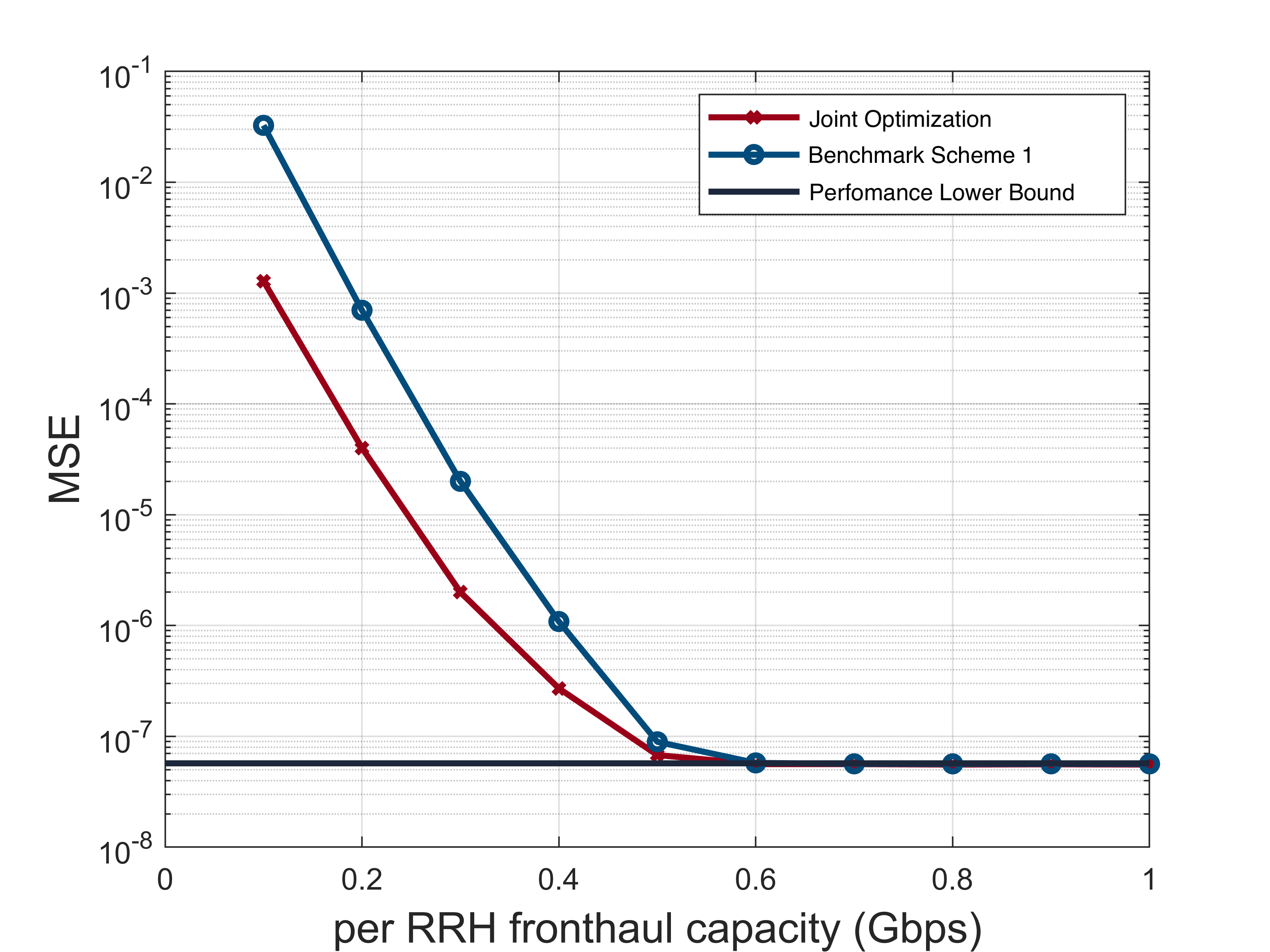}
	\caption{RRH fronthaul capacity versus the MSE of AirComp}
	\label{f1}
	\end{subfigure}
	
	\begin{subfigure} 
	\centering 
	\includegraphics[width=9cm]{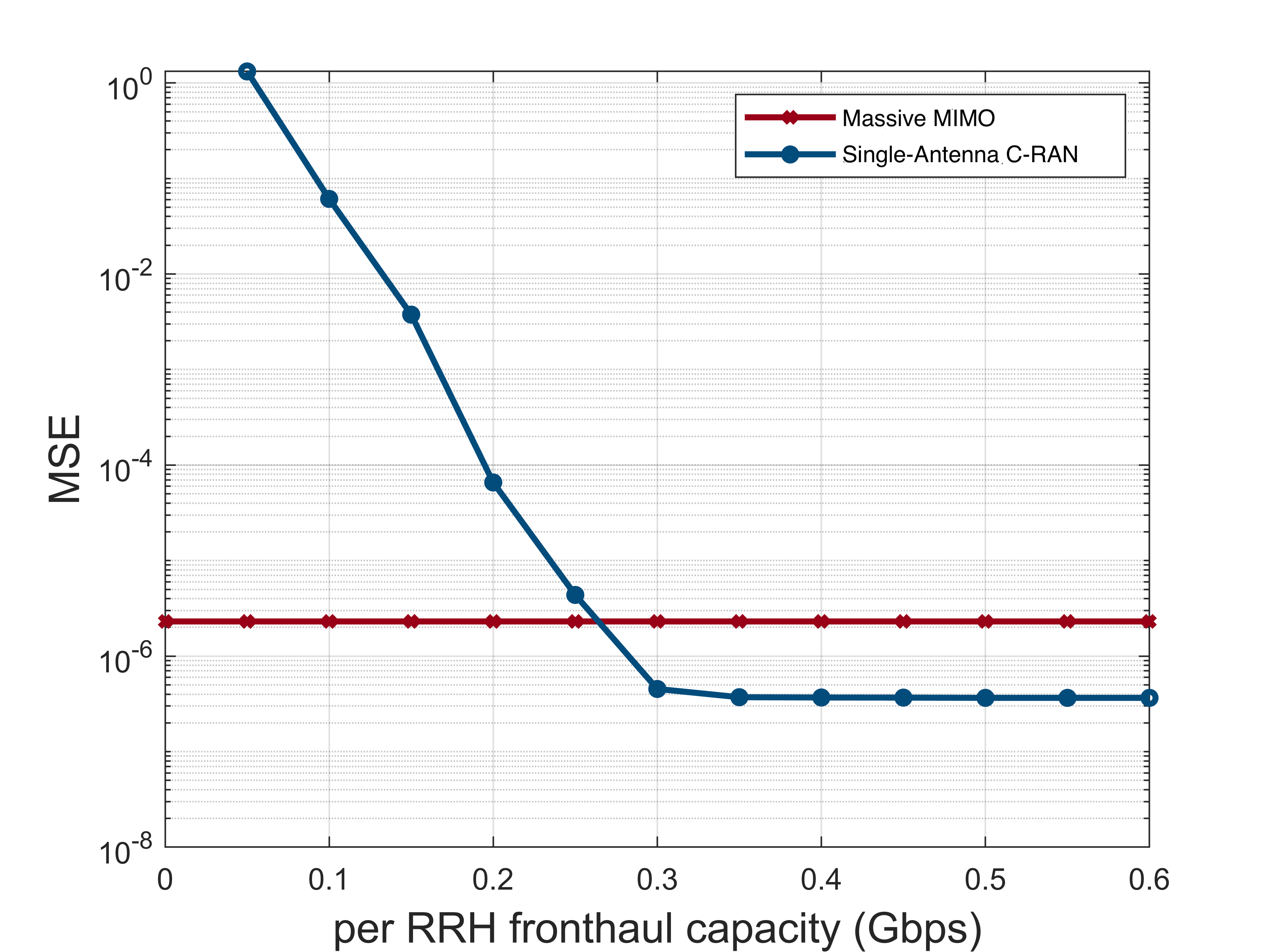}
	\caption{Performance comparison between single-antenna Cloud-RAN and massive MIMO }
	\label{f2}
	\end{subfigure}
\end{figure}

\section{Simulation Results}
In this section, we conduct numerical experiments to verify the feasibility of our proposed algorithm and compare the performance of AirComp in Cloud-RAN system with that in massive MIMO. We consider a 2-dimension coordinate system with $N_A$ RRHs and $N_W$ devices randomly distributed in a circle area of radius $ R = 500$ m. The bandwidth of the wireless channel is $B = 10$ MHz. We assume Rayleigh fading for all the considered channels and the pass loss model is formulated as 
\begin{equation}
	L(d) = T_0 \left( \frac{d}{d_0} \right)^{-\alpha},
\end{equation}
where $d$ is the distance of wireless link, $T_0$ is the reference pass loss corresponding to $d_0 = 1$ m and $\alpha$ denotes the pass loss exponent. In our numerical experiments, we set $T_0 = 30$ dB and $\alpha = 3$ for the wireless link between RRHs and devices. The maximum transmit power for each device is set to be $23$ dBm, the power spectral density of the background noise and the noise figure at each RRH is set to be $-169$ dBm and $7$ dB. Specifically, the channel coefficients are given by $\mathbf{h_{i,k}} = \sqrt{L(d_{i,k})\gamma^d}$, where $\gamma^d \sim \mathcal{CN}(0,\mathbf{I})$ and $d_{i,k}$ denotes the distance between device $k$ and RRH $i$. Furthermore, we assume that all the RRHs have the identical fronthaul capacity, i.e., $\bar T_i = T, \forall i$.

\subsection{Performance Gain of Quantization Bits Allocation at each RRH}
First, we show the performance gain obtained by quantization bits allocation at each RRH under the setting: $N_W = 15, N_A = 3, M = 8$. In addition to our proposed optimization algorithm, we consider a benchmark scheme and a performance lower bound (optimized quantization bits allocation and power control with setting $T$ to be infinite) as follows for comparision:

\begin{enumerate}
	\item \textbf{Benchmark Scheme 1 : Optimized power control with equal quantization bits allocation.} We consider that each RRH equally allocates its limited fronthaul capacity to all the signals their antennas received i.e., $C_{i,m} = \lfloor{\bar T_i/(2BM)}\rfloor,\forall i,m$. The BBU computes the transmit beamforming and receive beamforming vector by using algorithm \ref{overall algorithm} and not executing its Step 6-7.
	\item \textbf{Performance Lower Bound : Optimized power control with equal quantization bits allocation.} The performance lower bound can be easily obtained by Algorithm \ref{overall algorithm} by setting $T \rightarrow \infty$. At each RRH, the quantization noise power of each antenna goes to 0, i.e., $\omega_{i,m} \rightarrow 0, \forall i,m$.
\end{enumerate}

We show in Fig. \ref{f1} the MSE of AirComp versus different fronthaul capacity at each RRH. The MSE decreases significantly as the fronthaul capacity at each RRH increases and finally achieves the performance lower bound when the fronthaul capacity is large enough, which indicates larger fronthaul capacity brings better performance for AirComp in Cloud-RAN system. This is because as the fronthaul capacity increases, more bits can be utilized to quantize signal at each antenna. In this way, the quantization noise decreases. Furthermore, we can observe that as the fronthaul capacity is moderately small, our proposed joint optimization of quantization bits and transmit power enjoys a performance gain compared to the benchmark scheme.

\subsection{Comparison Between Cloud-RAN and Massive MIMO}
In this subsection, we compare the performance of AirComp in Cloud-RAN and massive MIMO architecture. To guarantee the fairness of our comparison, we fix the number of total antennas to be deployed. Next, in order to show the advantage of Cloud-RAN system, we study the following two cases: massive MIMO system with all of the antennas being deployed at the BS (which is located at origin of the given region), while for single-antenna Cloud-RAN, the RRHs are randomly located in the given region. Fig. \ref{f2} shows that MSE achieved in Cloud-RAN system is lower than the MSE achieved in massive MIMO system when appropriate number of fronthaul capacity is deployed at each RRH. However, if the fronthaul capacity deployed at each RRH is too small, the performance of AirComp in Cloud-RAN is inferior to massive MIMO system due to the large quantization noise. This result indicates that, our proposed architecture   for AirComp enjoys significant densification gain in reducing MSE because the RRHs are much closer to the devices compared with conventional massive MIMO architecture.

\section{Conclusions}
In this paper, we proposed to leverage the Cloud-RAN architecture to boost the performance for AirComp, thereby achieving accurate and ultra-fast data aggregation. To reduce the path loss of channels and thus provide reliable wireless connectivity to a large number of devices, we developed the Cloud-RAN architecture for AirComp. Then we formulated the optimization problem of joint devices' transmit beamforming and RRHs' quantization bits allocation and BBU's receive beamforming to minimize the MSE of AirComp. By applying our proposed alternating optimization method to solve this problem, the numerical results showed that our proposed approach can obtain the densification gain due to the massive deployment of RRHs compared to the massive MIMO system, and achieve better performance in reducing MSE of AirComp.

\bibliographystyle{IEEEtran}
\bibliography{refs} 

\end{document}